\documentclass[prd, twocolumn, nofootinbib]{revtex4}

\usepackage{epsfig}

\usepackage{color}
\definecolor{indigo}{rgb}{0.447,0.129,0.737}
\definecolor{brown}{rgb}{0.737,0.561,0.561}

\newcommand{\beq}{\begin{equation}}
\newcommand{\eeq}{\end{equation}}
\newcommand{\beqa}{\begin{eqnarray}}
\newcommand{\eeqa}{\end{eqnarray}}
\newcommand{\om}{\Omega_m}
\newcommand{\ome}{\Omega_e}
\newcommand{\omde}{\Omega_{DE}}
\newcommand{\dls}{d_{\rm lss}}

\newcommand{\ls}{\mathrel{\raise0.27ex\hbox{$<$}\kern-0.70em \lower0.71ex\hbox{{
$\scriptstyle \sim$}}}}

\begin{document} 

\title{Shifting the Universe: Early Dark Energy and Standard Rulers} 
\author{Eric V. Linder$^1$ and Georg Robbers$^2$}
\affiliation{$^1$Berkeley Lab \& University of California, Berkeley, CA 94720, 
USA \\ 
$^2$Institut f\"ur Theoretische Physik, Philosophenweg 16,
D-69120 Heidelberg, Germany}

\date{\today}

\begin{abstract} 
The presence of dark energy at high redshift influences both the cosmic 
sound horizon and the distance to last scattering of the cosmic microwave 
background.  We demonstrate that through the degeneracy in their ratio, 
early dark energy can lie hidden in the CMB temperature and 
polarization spectra, leading to an unrecognized shift in the sound 
horizon.  If the sound horizon is then used as a standard ruler, as in 
baryon acoustic oscillations, then the derived cosmological parameters 
can be nontrivially biased.  Fitting for the absolute ruler scale (just 
as supernovae must be fit for the absolute candle magnitude) removes 
the bias but decreases the leverage of the BAO technique by a factor 2. 
\end{abstract} 


\maketitle

\section{Introduction \label{sec:intro}}

The acceleration of the cosmic expansion occurs in the relatively recent 
universe, at redshifts $z<1$ when the universe is several billion years old. 
However, the origin of the dark energy responsible for the acceleration 
is unknown and we have no guarantee that it is of purely recent origin.  
Early dark 
energy models can be attractive both theoretically and observationally, 
and should not be neglected when interpreting cosmological data lest the 
conclusions be biased.  Here we consider the role that early dark energy 
(EDE) plays in high redshift observations of the cosmic microwave 
background (CMB) and baryon acoustic oscillations (BAO), and the use of 
the baryon-photon fluid sound horizon as a standard ruler. 

In fact, for any cosmological probe tied to 
high redshift one must know the expansion history, i.e.\ the dark energy 
properties, over the whole range or even the low redshift conclusions 
will be biased (see, e.g., \cite{pca}).  One cannot assume the dark 
energy ``fades away'': one must 
allow for the possibility of its early existence and test for this. 
Recognition of the impact of early dark energy is crucial for proper 
implementation of the current CMB shift parameter and BAO data constraints 
in cosmological parameter fitting, as illustrated in \cite{wright,kowalski} 
for example.  The effects on growth of structure are even more dramatic; 
see for example \cite{bartel,linearly}. 

CMB measurements are quite precise and the sound horizon is robust against 
many cosmological modifications, as shown by \cite{eiswh}, but they also 
identified that the early expansion history represents a loophole in this. 
Early dark energy contributing a fraction $\ome$ to the total energy 
density shifts the sound horizon by a factor $(1-\ome)^{1/2}$ 
\cite{dst}.  By itself this would be precisely measured in CMB data; 
however the distance to CMB last scattering changes as well and could 
compensate such that both shifts went undetected \cite{chall}.  

In this paper we examine the effects on CMB and BAO data more closely. 
Describing early dark energy in \S\ref{sec:early}, we investigate 
in \S\ref{sec:geom} first the geometric degeneracy between the sound horizon 
and the distance to last scattering and then carry out a Markov Chain 
Monte Carlo analysis of the CMB power spectra to give a detailed comparison 
of early and standard fading dark energy cosmologies.  In \S\ref{sec:bao} 
we examine the consequences of unrecognized shifts in the cosmology for 
the BAO method employing the biased standard ruler.

\section{Early dark energy model \label{sec:early}}

Early dark energy has motivations on both the theoretical and observational 
fronts.  Tracking quintessence \cite{tracker} attempts to alleviate the 
coincidence problem of the cosmological constant by considering a dynamical 
scalar field with a potential of the form that brings the field evolution 
onto an attractor trajectory, greatly enlarging the range of initial 
conditions that can deliver a present dark energy density near 70\% of 
the critical energy density.  This has the consequence that the energy 
density is a nonnegligible fraction of the matter density during the matter 
dominated era, including the recombination epoch.  Dilaton fields giving 
rise to an exponential potential 
\cite{wett88}, appearing in many particle physics models, have such a form 
and indeed the dark energy density traces the dominant energy density as 
a constant fraction.  

In addition to persistent, early dark energy some possible sources for 
deviation in the sound horizon paradigm include a post-nucleosynthesis 
acceleration as in stochastic dark energy models \cite{dodel,baren} or 
nonminimal couplings such as in scalar-tensor \cite{scatens} or coupled 
dark energy theories \cite{amen}.  However we concentrate on EDE as the 
simplest and directly motivated model. 

Constraints from primordial nucleosynthesis and from the CMB limit the 
early dark energy fraction to be small, less than about 4\% \cite{drw}, 
but not completely negligible.  Indeed, EDE at the 1\% level can have 
significant effects on formation of massive structures \cite{bartel} 
and may help explain the early onset of star and galaxy formation 
as well as the high level of Sunyaev-Zel'dovich effect 
contribution to the high multipoles of the CMB temperature power spectrum 
\cite{silk07061340}.  Note that in contrast, a cosmological constant has 
a fractional energy density at the $10^{-9}$ level at $z\approx10^3$. 

To capture the physical effects of dark energy at both recent and early 
epochs we adopt the form \cite{dr} 
\beq 
\omde(a)=\frac{\omde-\ome\,(1-a^{-3w_0})}{\omde+\om a^{3w_0}} 
+\ome\,(1-a^{-3w_0}) \label{eq:heid}
\eeq
for the dark energy density as a function of scale factor $a=1/(1+z)$.  
Here $\omde$ is the present dark energy density,
$\ome$ is the asymptotic early
dark energy density, $w_0$ is the present dark energy equation of state,
or pressure to energy density ratio, and $\om\,(=1-\omde)$ is the present
matter density.

\begin{figure*}[!htb]
	\centering
	\includegraphics[width=0.9\textwidth]{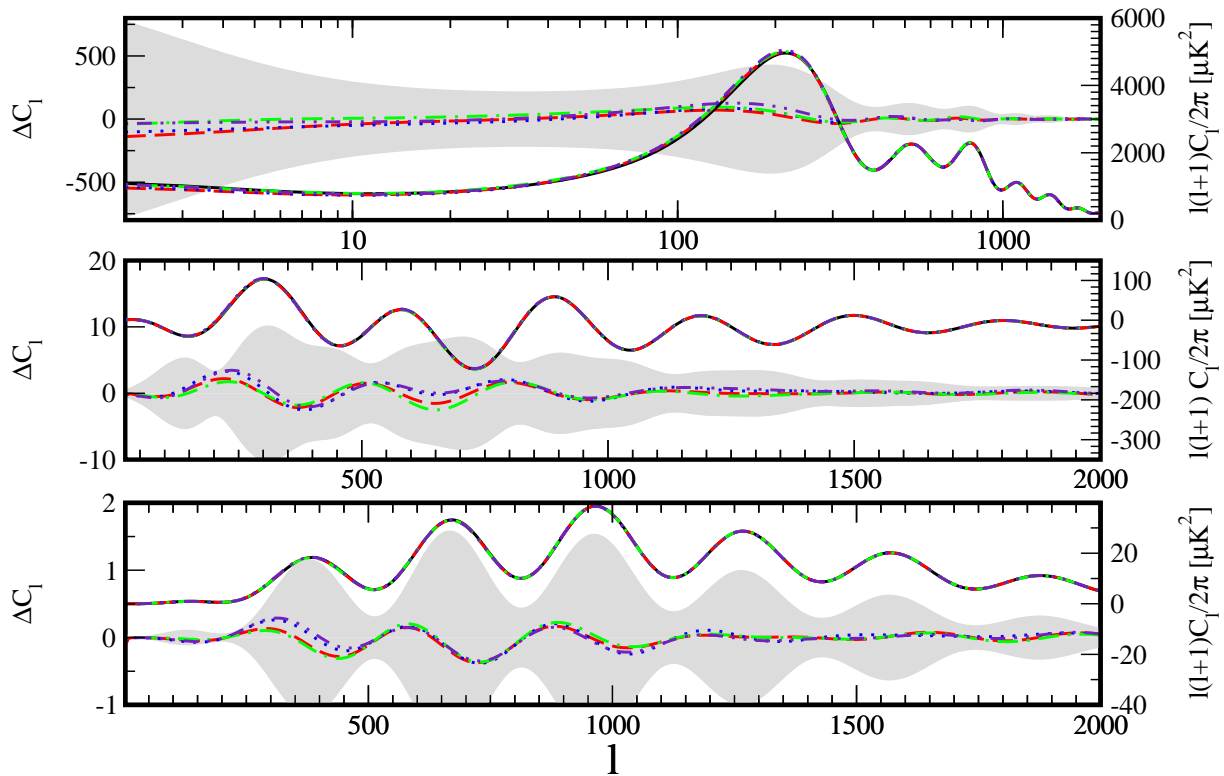}
	\caption{CMB power spectra for TT (top panel), TE (middle), 
and EE (bottom) modes of different $\ome=0.03$ EDE models (listed in 
Table~\ref{tab:cllegend}) are plotted using the right axis scale.  
Their differences relative to the fiducial $(w_0=-0.95, w_a=-0.1)$ 
model are shown using the left axis scale, and lie within the grey 
shading surrounding the zero difference level that shows the cosmic 
variance limit. 
}
	\label{fig:TTwa}
\end{figure*}

\section{CMB Power Spectra \label{sec:geom}}

The sound horizon of the coupled photon-baryon fluid plays a critical 
role in the physics of density perturbations in both the photons (which 
translate into the temperature perturbations seen through the CMB 
temperature power spectrum) and baryons (which appear as baryon acoustic 
oscillations, seen through the statistical pattern of galaxy positions, 
today on $100\,h^{-1}$Mpc scales).  The sound horizon $s$ is given by 
\beq 
s=\int_{z_{dec}}^\infty dz\frac{c_s}{H(z)}, 
\eeq 
where $z_{dec}$ is the redshift of decoupling between the photons and 
baryons, $c_s$ is the sound speed of the fluid, depending on the baryon 
and photon densities, and $H(z)$ is the Hubble parameter, depending on 
the energy densities of all components, not just baryons and photons. 

As mentioned, in $\Lambda$CDM the cosmological constant contributes less 
than $10^{-9}$ of the dominant term in $H(z)$, and even for a dark energy 
equation of state $w=-2/3$ the contribution is of order $10^{-6}$.  
The Hubble parameter for a universe with early dark energy in addition 
to the usual (matter and radiation) components is given by 
\beqa 
H^2(z)&=&\frac{8\pi}{3}[\rho_{\rm noEDE}(z)+\rho_{\rm EDE}(z)] \\ 
&=&\frac{8\pi}{3}
\rho_{\rm noEDE}(z)+H^2(z)\Omega_{\rm EDE}(z) \\ 
&=&\frac{8\pi}{3}\frac{\rho_{\rm noEDE}(z)}{1-\Omega_{\rm EDE}(z)}\,. 
\eeqa 
Thus we see that for EDE scaling as a constant fraction of the energy 
density that at high redshift $H(z)\sim (1-\ome)^{-1/2}$ \cite{dor00}. 
Holding $z_{dec}$ fixed (it depends predominantly on atomic 
physics and is quite robust to changes in dark energy) and $c_s$ fixed 
(it involves baryonic physics), then we immediately see that 
\beq 
\frac{s_{\rm EDE}}{s_{\rm noEDE}}\sim (1-\ome)^{1/2}\,. \label{eq:sede}
\eeq
Thus, early dark energy shifts the sound horizon.

Furthermore, EDE shifts distances, for example the conformal distance 
\beq 
d(z)=\int_0^z \frac{dz'}{H(z')}, 
\eeq 
but the crucial point is that because the dark energy contributes 
different amounts to $H(z)$ at different redshifts, the shifts are not 
uniform but vary with redshift.  Thus, the measured distance ratio, or 
angular scale, of the distance relative to the sound horizon scale, is 
preserved for only one particular distance, depending on the 
early dark energy density and other cosmological parameters.  

For some parameter values, 
the ratio of distance to CMB last scattering relative to the sound horizon, 
$d_{\rm lss}/s$, is conserved, leaving untouched the geometric factor 
determining the measured location of the CMB acoustic peaks.  This has 
two implications: one cannot distinguish early dark energy from standard 
fading dark energy in this case from the CMB acoustic scale, and then one 
unavoidably obtains a shift in the ratio of sound horizon to distances 
at other redshifts -- precisely what enters BAO.  Without specifically 
fitting for the possibility of EDE therefore, the shift goes unrecognized, 
effectively miscalibrating the distances, and pointing to a spurious 
cosmology.  

To calibrate the standard ruler of the sound horizon for BAO cosmology, 
one must therefore either discern signatures of EDE in the CMB even with 
$\dls/s$ matched, or employ independent observations, such as the 
amplitude of growth and nonlinear structure.  We 
concentrate here on the former, since precision measurements of growth 
are not as advanced. 

We now investigate whether one can robustly recognize models with EDE 
distinctly from fading dark energy cosmologies by examining the CMB power 
spectra for those models even when the last scattering distance ratio 
is matched.  For example, the CMB peak structure is not a purely
periodic oscillation but is driven by the gravitational potentials. 
Since EDE is essentially unclustered at these early times, it does not 
participate in the driving as dark matter does, giving an early Sachs-Wolfe
effect that alters the rise to the first 
peak\footnote{Note that attributing the effect of EDE to the early 
Sachs-Wolfe effect, as opposed to say the Doppler rise from the 
perturbation velocities, is a gauge dependent statement.  See \cite{araahu} 
for a review of the various physical effects that contribute to the CMB
power spectrum.}.
However, this  might in turn be compensated by tilting the scalar index, 
which then has further physical effects that might be compensated by 
changing other parameters.  

Figure~\ref{fig:TTwa} demonstrates just how exactly models with EDE can
reproduce the CMB power spectra of a fiducial fading dark energy model. 
For Fig.~\ref{fig:TTwa}, we chose the CMB power spectra of a 
$w_0=-0.95$, $w_a=-0.1$ model without EDE as the fiducial spectra, and then 
used the Boltzmann code cmbeasy \cite{cmbeasy, analyzethis}
to find EDE models with $\ome=0.03$,  $w_0=-0.95$
that match the fiducial spectra\footnote{We chose $w_a=-0.1$ to match the 
CMB distance ratio of the EDE model with $\ome=0.03$ but all other 
parameters fixed.}.    
To find matching spectra, we scanned the parameter space
via Markov Chain Monte Carlo (MCMC) runs,
varying the physical baryon density $\Omega_b h^2$, physical matter density 
$\om h^2$, Hubble constant $h$, optical depth $\tau$, 
power spectrum amplitude $A_s e^{-2\tau}$, and the scalar spectral 
index $n_s$ of the EDE models. 

\begin{table}
\begin{tabular}{|ccp{0.4\columnwidth}|} \hline
fiducial & {\huge------} & $w_0=-0.95$, $w_a=-0.1$, no EDE \\ \hline
A &{\color{red} \huge  \bf -- --} & matching $2<C_\ell^{TT}<2000$ \\
\hline
B & {\color{blue}  \bf \huge $\cdots$}& $\om=0.28$, matching
$2<C_\ell^{TT}<2000$ \\ \hline
C & {\color{green} \bf \huge $\cdot$ -- $\cdot$} &
matching $2<C_\ell^{TT,EE}<2000$\\ \hline
D &{\color{indigo} \bf \huge -- $\cdot \; \cdot$} &$\om=0.28$,
matching $2<C_\ell^{TT,EE}<2000$\\ \hline
\end{tabular}
\caption{Fitting requirements for the models shown
             in Fig.~\ref{fig:TTwa}. }
\label{tab:cllegend}
\end{table}

The examples shown are from Monte-Carlo chains trying to
match the TT-spectra in the range $2 < \ell < 2000$ (model A), plus
making the match more challenging by requiring exact agreement on
$\om$ (model B), or matching both TT and EE-spectra (model C), plus
exact $\om$ (model D) (also see Table~\ref{tab:cllegend}).  
As seen, the CMB power spectra of models containing substantial early 
dark energy, here $\ome=0.03$, can mimic a cosmology lacking 
early dark energy.  That is to say, the shift in the sound horizon and 
subsequent miscalibration of the standard ruler can be hidden.  

For all matching requirements, the comparison precision assumed was 
for a experiment covering a fraction of 81\% of the sky, limited only 
by cosmic variance up to $\ell = 2000$, which Planck will approach for 
the TT power spectrum.  
The overall $\chi^2$ for the TT-spectrum of model A relative to the 
fiducial model, for six parameters over nearly 2000 degrees of freedom, 
is 14.  Note for comparison that WMAP 5 year data gives a $\chi^2$ of 
$\sim1000$ relative to their best fit $\Lambda$CDM model for unbinned 
multipoles over the range $\ell=33-1000$ \cite{WMAPangpow}.  
Their reduced $\chi^2$ with binned multipoles is 1.04; using the 
same binning the EDE model has a reduced $\chi^2$ of 0.3. 
Moreover, we have considered only a limited set of potential degeneracies, 
for example holding $w_0$ fixed and not including running or spatial 
curvature, so further 
degeneracies exist that could aid in hiding early dark energy.  

Two possible 
routes to breaking the degeneracies that prevent recognition of EDE 
are use of the B-mode spectrum, whose lensing component measures growth 
of structure, affected by EDE as mentioned above (although note that 
the ISW effect already included also probes growth), and precision measurement 
of the Hubble constant $H_0$, that shifts to partially compensate for 
the sound horizon alteration.  

It is also interesting to note the implication that the physical matter 
density $\om h^2$ 
entering predictions for dark matter direct detection signals may differ 
from the value from standard CMB constraints.  Since the nonrelativistic 
matter density needed to fit the CMB data gains an additional contribution 
from EDE, i.e.\ is $(\om+\ome)h^2$ if EDE had $w=0$ (but in fact $w$ 
around recombination is often slightly positive), 
then the quantity $\om h^2$ may differ from what the standard analysis 
might suggest. 

The MCMC results presented here demonstrate that the 
presence of EDE (even with the rather large value of $\ome=0.03$) may 
not be cleanly 
distinguished from standard fading dark energy, even by cosmic variance 
limited CMB experiments similar to Planck.  In the next section we 
investigate the implications if such a miscalibration were to occur.

\section{Nonstandard Ruler \label{sec:bao}}

\begin{figure}
	\centering
	\includegraphics[angle=270, bb=29 0 589 835,
width=0.45\textwidth]{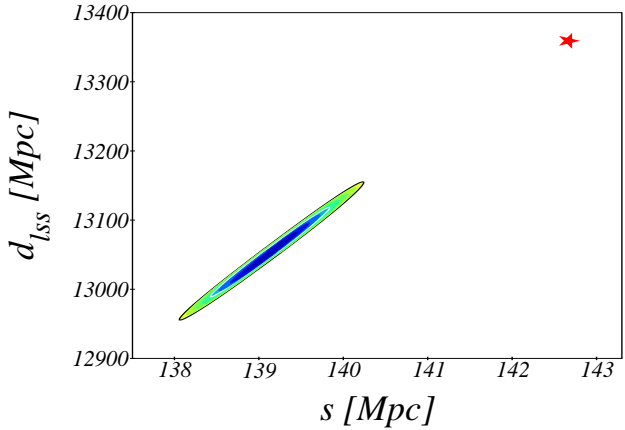}
       \caption{Likelihood contour at 95.4\% cl in the $s-\dls$ plane for EDE
         cosmologies with $\ome=0.03$ that match the CMB 
         temperature power spectrum of the
         fiducial \mbox{$(w_0=-0.95, w_a=-0.1)$} model without early dark
         energy.  Note that the matching 
almost automatically preserves the geometric shift factor, i.e.\ the 
slope $\dls/s$, but not the individual distance scales.  
         The red star marks $s$  and $\dls$ of the fiducial model; the 
shift in the standard ruler amounts to 2.5\%.
       }
	\label{fig:like2Dwasd}
\end{figure}

Cosmological probes such as BAO that rely on the sound horizon (or any
other early universe quantity) as a standard ruler will be
miscalibrated in the presence of unrecognized EDE.  Both the sound horizon 
$s$ and distances $d$ to various redshifts are altered.  It is the ratio 
of these that provides cosmological tests.  Figure~\ref{fig:like2Dwasd} 
illustrates the dramatic shifts in each quantity; note that the shift 
exceeds that in Eq.~(\ref{eq:sede}) because other parameters change 
between the fit and fiducial models.  The figure also demonstrates how 
the shifts can be hidden through preservation of the CMB acoustic peak 
scale given by the ratio of the distance to last scattering to the sound 
horizon scale, $\dls/s$.

The effect of the miscalibration of the sound horizon standard ruler
in the distance ratio varies with redshift and so can mimic changes in the 
cosmology.  We examine the propagation of this miscalibration in BAO 
analysis into the biasing of cosmological parameter estimation.  
In particular, we consider the effect on the derived dark energy 
equation of state parameters of the form 
\beq 
w(z)=w_0+w_a\,(1-a). \label{eq:wa}
\eeq 

Within the Fisher information formalism, a miscalibration in the measured 
quantities $O_k$ propagate into the cosmological parameters $p_i$ as 
\beq 
\delta p_i=(F^{-1})_{ij} \sum_k \Delta O_k \frac{\partial O_k}{\partial p_j} 
\frac{1}{\sigma^2(O_k)}, \label{eq:bias}
\eeq 
approximating measurement errors in the observables as uncorrelated. 
For BAO the ``observables'' are taken to be the ratio of the angular diameter 
distance at some redshift $z_k$ to the sound horizon, $\tilde d=d/s$, and the 
ratio of the proper distance interval (essentially $H^{-1}(z_k)$ for 
infinitesimal intervals) to the sound horizon, $\tilde H^{-1}=H^{-1}/s$.  
These arise from measurement of the transverse and radial BAO modes, 
respectively. 

\begin{figure}[!htb] 
  \begin{center}
     \psfig{file=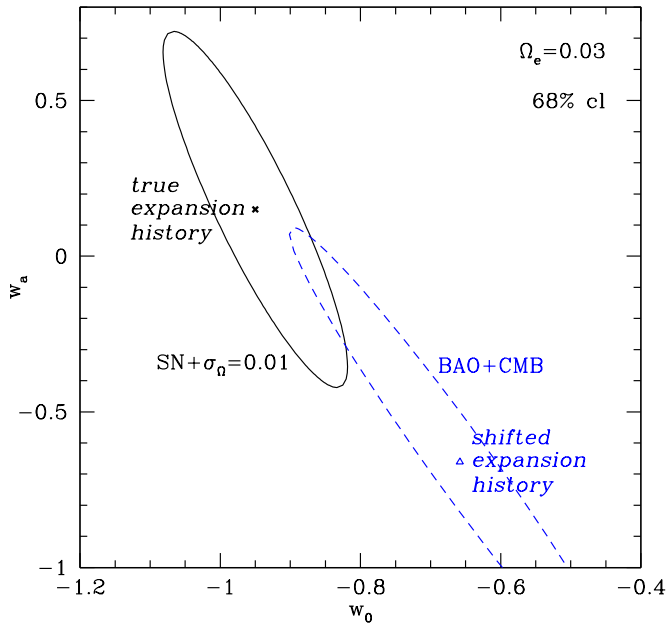,width=3.5in}
  \caption{Early dark energy density $\ome$ can cause miscalibration of 
the standard ruler used by the baryon acoustic oscillation probe.  This 
shifts the dark energy properties derived with BAO+CMB data from their 
true values (black x) to those marked by the blue triangle, with the dashed 
confidence contour.  Distance probes such as supernovae that do not rely 
on the standard ruler recover the true expansion history. 
  }
  \label{fig:boscbias}
  \end{center}
\end{figure}

Early dark energy shifts the sound horizon, the distances, and the distance 
intervals, and from these generates the offsets $\Delta O_k$.  
The bias on the derived cosmology due to the miscalibration is illustrated 
in Fig.~\ref{fig:boscbias}.  Here we take a true cosmology with $\ome=0.03$, 
$\om=0.28$, $w_0=-0.95$ and consider the results of a BAO experiment 
achieving 1\% precision on both $\tilde d$ and $\tilde H^{-1}$ at 
redshifts $z=0.4$, 0.6, 0.8, 1.  Note that this is quite optimistic, being 
beyond the statistical power of the proposed BOSS and WFMOS surveys even 
without accounting for systematics.  We also include CMB measurements of 
the quality of Planck.  The combination BAO+CMB represents the cosmological 
probes tied to high redshift. 

The interplay between the shifts in the sound horizon and the distances 
acts to bias $w_0$ by a substantial $\Delta w_0=0.29$, and mimic an evolution 
$w_a=-0.66$.  More precise BAO measurements would lead to even greater 
relative bias.  Note that in other models for EDE, 
such as the mocker model considered in \cite{chall}, BAO can give even 
larger biases for the same amount of EDE.

Observations that are not calibrated off the high redshift universe do 
not suffer this miscalibration bias.  Figure~\ref{fig:boscbias} shows 
the confidence region for luminosity distance measurements through Type Ia 
supernovae (SN), at 1\% accuracy over the range $z=0-1.7$ as from the 
proposed SNAP mission, combined with a prior on the matter density 
mimicking what might be delivered by its gravitational weak lensing 
measurements of distance ratios.  These measurements recover in an 
unbiased manner the true expansion history over this redshift range.  
That is, the luminosity distance predictions of the best fit 
$(w_0,w_a)=(-0.95,0.15)$ model agree with the true EDE cosmology to 
better than 0.02\%. (The expansion history of an EDE model at $z\lesssim2$ 
closely follows a standard dark energy model with the same $w_0$ and 
with an effective $w_a\approx 5\,\ome$.) 

In order for BAO to recover the true cosmology, the technique must allow 
for the possibility of EDE; one must include a calibration parameter to 
standardize the ruler, i.e.\ a new parameter $s$ or $\delta s$.  This is 
exactly analogous to the ${\mathcal M}$ parameter for SN that allows for 
the possibility that the low redshift calibration is not perfect, i.e.\ 
SN analysis does not assume they are known standard candles but fits for 
the absolute luminosity.  When we 
include such a parameter in the fit, the bias goes away while the 
parameter estimation uncertainties increase of course.  

The necessary 
presence of a standard ruler calibration parameter, call it ${\mathcal S}$, 
leads to an increase in the $w_0$-$w_a$ contour area, and equivalent 
decrease in the area ``figure of merit'' by a factor 2.3.  (To decrease 
appreciably the area blowup would require prior constraint on 
${\mathcal S}$ to better than 0.5\%.)  Since we do not 
know a priori whether the high redshift universe is shifted by EDE, 
or by other physics mentioned in \S\ref{sec:early}, neglecting 
${\mathcal S}$ for BAO is as improper as neglecting 
${\mathcal M}$ for supernova 
standard candle calibration.  Without the need to fit for the low 
redshift calibration ${\mathcal M}$, SN would enjoy an improvement in 
``figure of merit'' by a factor 1.9, similar to the 2.3 that BAO has when 
neglecting the high redshift calibration ${\mathcal S}$.  In summary, 
if calibration is ignored then BAO may deliver biased contours 
as in Fig.~\ref{fig:boscbias}, while if calibration is implemented as 
described then the BAO contours are centered on the true cosmology but 
increase by $\sim2.3$ in area.

\section{Conclusion \label{sec:concl}}

Early dark energy has possible motivations both theoretically and 
observationally, but more than that we simply should not {\it assume\/} 
dark energy is a purely recent phenomenon.  Such an assumption can 
significantly bias our understanding of the cosmological model even 
at low redshift.  We have shown that recognizing early dark energy 
in the cosmic microwave background is not trivial, even if 
it significantly affects the size of the sound horizon. 

A level of $\ome>0.03$ would likely provide a recognizable signal in 
advanced CMB data.  However, lesser amounts of early dark energy capable 
of shifting the universe by $\sim1\%$ are more difficult to identify 
robustly. 

The sound horizon is a crucial quantity because it is employed as a 
standard ruler in the baryon acoustic oscillation technique for probing 
cosmology.  If this standard ruler is miscalibrated due to unrecognized 
early dark energy, then bias will enter into interpretation of 
BAO scale measurements.  The remedy is inclusion of 
a calibration parameter, just as exists for the supernova standard candle 
technique, which removes the bias but weakens the BAO constraints by a 
factor $\sim2$.  

Probing, rather than assuming the lack of, early dark energy 
offers an intriguing window on the new physics behind the accelerating 
universe.  Structure formation as probed through the first stars, 
cluster abundances and the Sunyaev-Zel'dovich effect, and weak gravitational 
lensing may hold important clues to early dark energy, while precision 
measurements of the Hubble constant can provide additional leverage for 
CMB constraints.  Understanding dark energy throughout the history of the 
universe is an exciting prospect.

\acknowledgments 

EL thanks Christof Wetterich for hospitality at ITP Heidelberg during 
the course of this project, and Michael Doran and Martin White for useful 
conversations. 
This work has been supported in part by the Director, Office of Science, 
Department of Energy under grant DE-AC02-05CH11231. 
GR acknowledges support by the Deutsche Forschungsgemeinschaft,
grant TRR33 ``The Dark Universe''.

\end{document}